# LOCx2, a Low-latency, Low-overhead, 2 × 5.12-Gbps Transmitter ASIC for the ATLAS Liquid Argon Calorimeter Trigger Upgrade

**Le Xiao,[a,b] Xiaoting Li,[a] Datao Gong,[b,*] Jinghong Chen,[c,g] Di Guo,[b,d] Huiqin He,[a,b,e] Suen Hou,[f] Guangming Huang,[a,*] Chonghan Liu,[b] Tiankuan Liu,[b] Xiangming Sun,[a] Ping-Kun Teng,[f] Bozorgmehr Vosooghi,[c] Annie C. Xiang,[b] Jingbo Ye,[b] Yang You,[g] Zhiheng Zuo [c]**

[a] *Department of Physics, Central China Normal University, Wuhan, Hubei 430079, P.R. China*

[b] *Department of Physics, Southern Methodist University, Dallas, TX 75275, USA*

[c] *Department of Electrical and Computer Engineering, University of Houston, Houston, TX 77004, USA*

[d] *State Key Laboratory of Particle Detection and Electronics, University of Science and Technology of China, Hefei, Anhui 230026, China*

[e] *Shenzhen Polytechnic, Shenzhen 518055, P.R. China*

[f] *Institute of Physics, Academia Sinica, Nangang 11529, Taipei, Taiwan*

[g] *Department of Electrical Engineering, Southern Methodist University, Dallas, TX 75275, USA*

*E-mail*: Guangming Huang (gmhuang@phy.ccnu.edu.cn) and Datao Gong (dgong@mail.smu.edu)

ABSTRACT: In this paper, we present the design and test results of LOCx2, a transmitter ASIC for the ATLAS Liquid Argon Calorimeter trigger upgrade. LOCx2 consists of two channels and each channel encodes ADC data with an overhead of 14.3% and transmits serial data at 5.12 Gbps with a latency of less than 27.2 ns. LOCx2 is fabricated with a commercial 0.25-µm Silicon-on-Sapphire CMOS technology and is packaged in a 100-pin QFN package. The power consumption of LOCx2 is about 843 mW.

[*] Corresponding author.

## Contents

## 1. Introduction

The Large Hadron Collider (LHC) will be upgraded in 2018-2019 to reach about three times the current luminosity. Since the current ATLAS trigger system cannot support such luminosity increase, a new trigger system is required. As part of the new trigger system, the ATLAS Liquid Argon Calorimeter (LAr) readout system is being developed to provide digital trigger signals for the trigger to improve background rejection. In the LAr Phase-I trigger upgrade [1], 34000 high-granularity signals will be sampled and digitized on 124 LAr Trigger Digitizer Boards (LTDBs), each handling up to 320 signals. Each LTDB will transmit all digitized data out of the front end via 40 optical fibers. The total payload rate of each LTDB is about 180 Gbps.

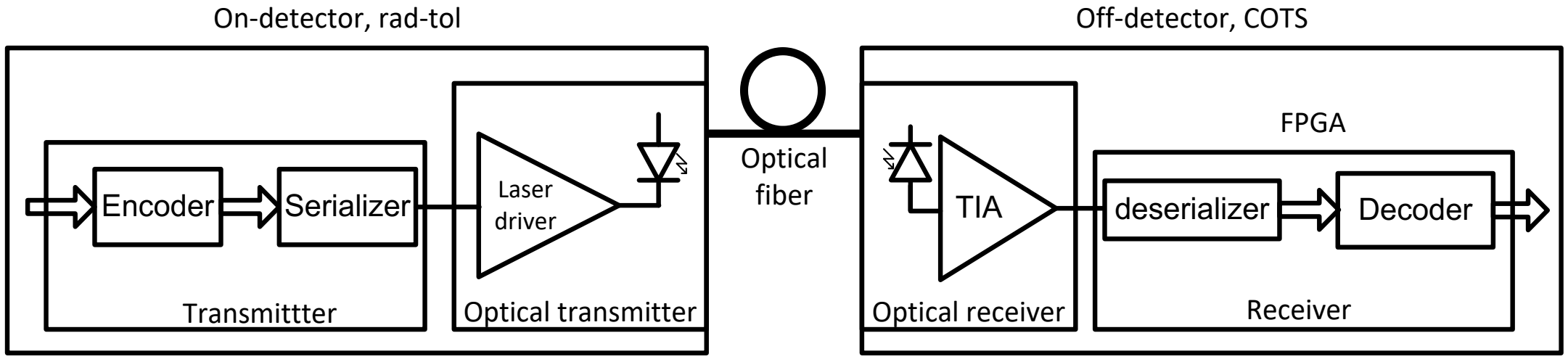

**Figure 1:** The block diagram of optical data transmission system.

Figure 1 shows a generic block diagram of optical data transmission system. The transmitter side includes an encoder, a serializer, and an optical transmitter. The transmitter, which integrates the encoder and the serializer, encodes the data coming from the upstream ADCs and provides a serial output. The optical transmitter converts the electrical signal to an optical signal. On the receiver side, an optical receiver converts the signal coming from the optical fiber into an electrical signal. The deserializer and the decoder then recover the original payload from the serial data and check if any error occurs in the data transmission.

The requirements of the ATLAS LAr Phase-I trigger upgrade on the transmitter are summarized as follows: First, since the transmitter is installed at the front end, no commercial transmitter can tolerate the harsh radiation environment [2-3]. Therefore, a transmitter Application-Specific Integrated Circuit (ASIC) is required. Second, since the transmitter is used in the trigger system, the latency of the transmitter must be small, requiring careful selection of the line code and a low-latency design of the transmitter. The latency budget of the optical link (not including the time passing through the optical fiber) is 150 ns. Third, the power budget of 100 mW per Gbps puts stringent constraint on the power consumption and the overhead of the transmitter.

In this paper, we present a transmitter ASIC, LOCx2, designed for the ATLAS LAr Phase-I trigger upgrade. LOCx2 consists of two channels. Each channel encodes the ADC data with an overhead of 14.3%, serializes the encoded data, and transmits the serial data at 5.12 Gbps. LOCx2 is fabricated with a commercial 0.25-μm Silicon-on-Sapphire CMOS technology that has been evaluated to be suitable for the ASIC development in front-end readout systems [4-6].

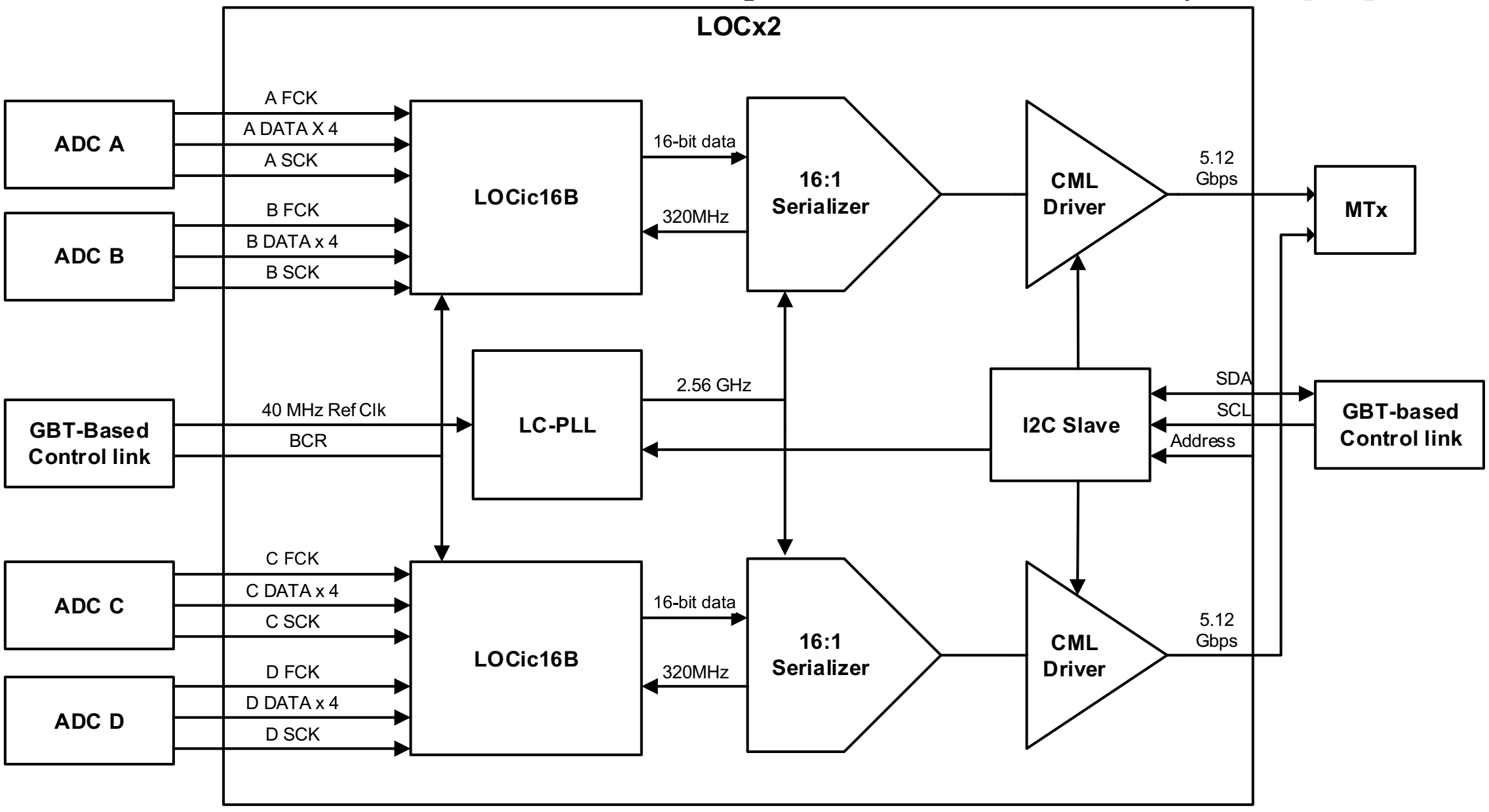

**Figure 2**: The block diagram of LOCx2.

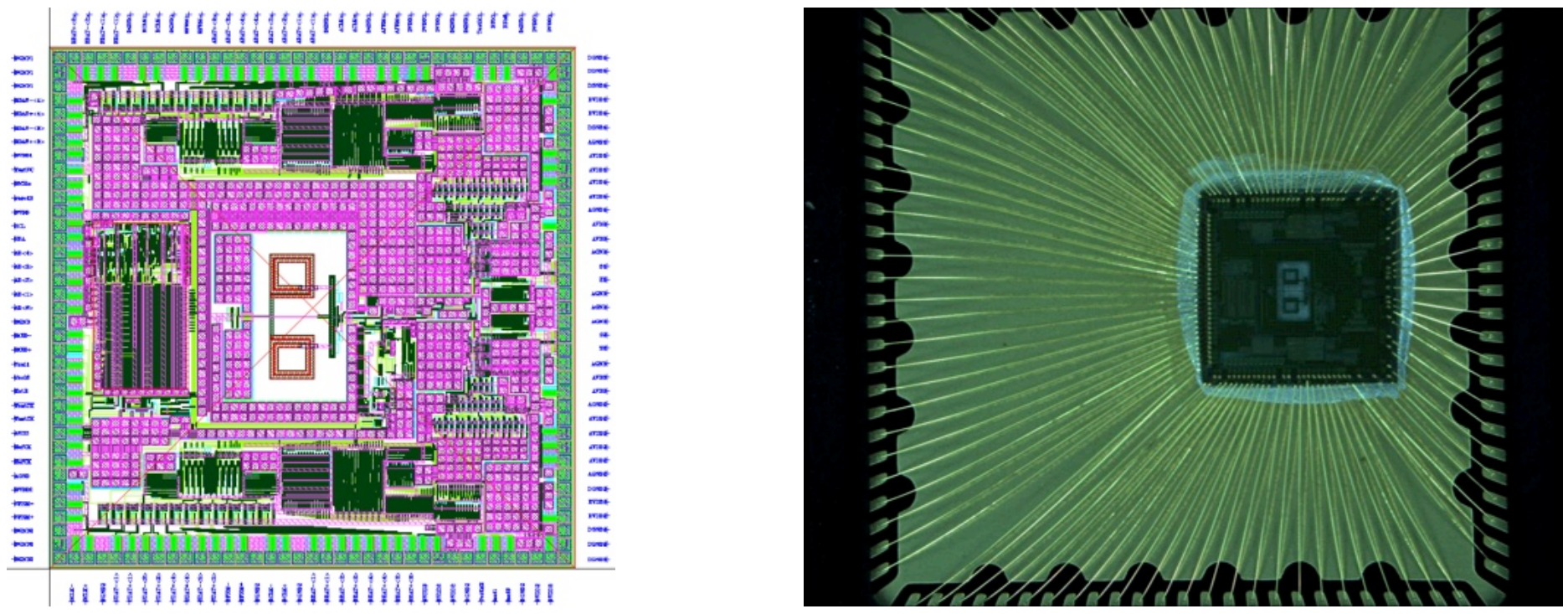

(a) (b)

**Figure 3:** The layout of LOCx2 (a) and the bonding micrograph of the QFN package (b).

## 2. The design of LOCx2

LOCx2 has the following functional blocks: two encoders, two 16:1 serializers, two Current Mode Logic (CML) line drivers, a shared Phase-Locked Loop (PLL) with an inductor-capacitor-tank-based Voltage Controlled Oscillator (LC-VCO), and a shared Inter-Integrated Circuit ($I^2C$) slave module. The block diagram of LOCx2, as well as its interfaces with the ADC ASIC [7], GBTX [8], and MTx [9], is shown in figure 2. LOCx2 has two serializers and each serializer takes the digitized data of eight analog signals from two ADC chips. LOCx2 requires a 40-MHz LHC reference clock and an LHC bunch-crossing reset (BCR), both provided by the GBT-based control link [10]. LOCx2 has 32-bit internal registers, which are configured via an $I^2C$ slave module. LOCx2 is packaged in a 100-pin QFN package. The whole chip layout and the bonding diagram of the QFN package are shown in figure 3.

### 2.1 The PLL

The block diagram of the PLL is shown in figure 4. The input reference clock is 40 MHz and the output clock is 2.56 GHz. The PLL is composed of a Phase-Frequency Detector (PFD), a current-programmable charge pump, a Low-Pass Filter (LPF), a divided-by-64 divider chain, and an LC-VCO. The PLL is optimized to operate at 2.56 GHz.

The LC-VCO has four tuning frequency bands selected via an $I^2C$ interface. The tuning range of the PLL is simulated to be from 1.86 GHz to 2.98 GHz at the nominal process corner and 55 ℃.The VCO phase noise is less than -105 dBc at 1 MHz offset from the carrier frequency, corresponding to a random jitter of less than 1 ps (RMS).

Both the bandwidth of the LPF and the current of the charge pump are programmable. The loop bandwidth of the PLL is programmable from 0.5 to 2.5 MHz. The LPF can be configured to be either a 3$^{rd}$- or a 2$^{nd}$-order filter. The 3$^{rd}$-order LPF can suppress the spur in the PLL. The 2$^{nd}$-order filter has larger phase margin than the 3$^{rd}$-order one. The 3$^{rd}$-order LPF has a phase margin of more than 45 degree and the 2$^{nd}$-order LPF has the phase margin of larger than 60 degree.

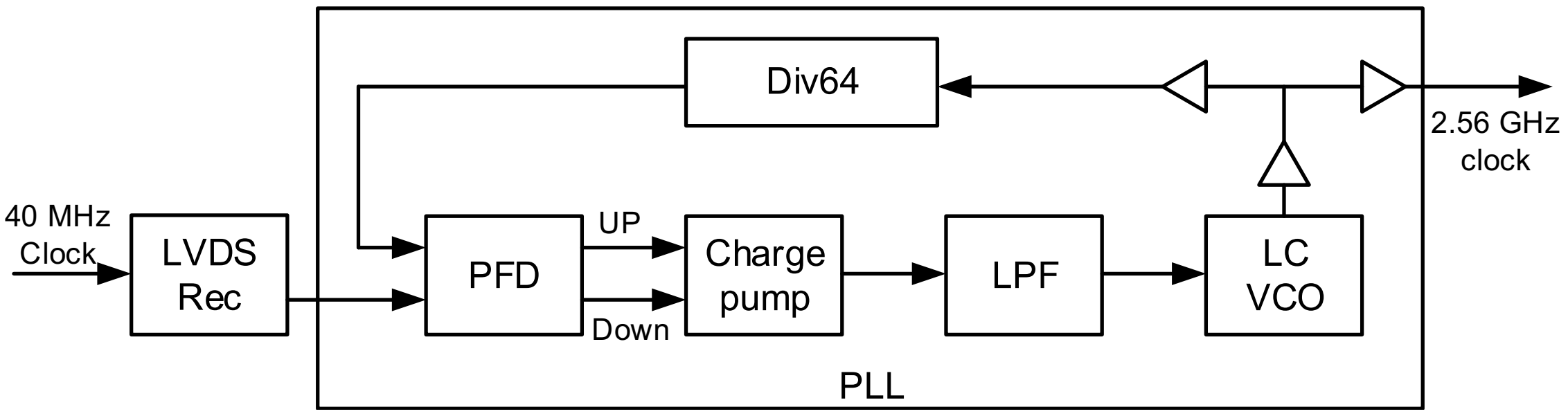

**Figure 4**: The block diagram of the PLL.

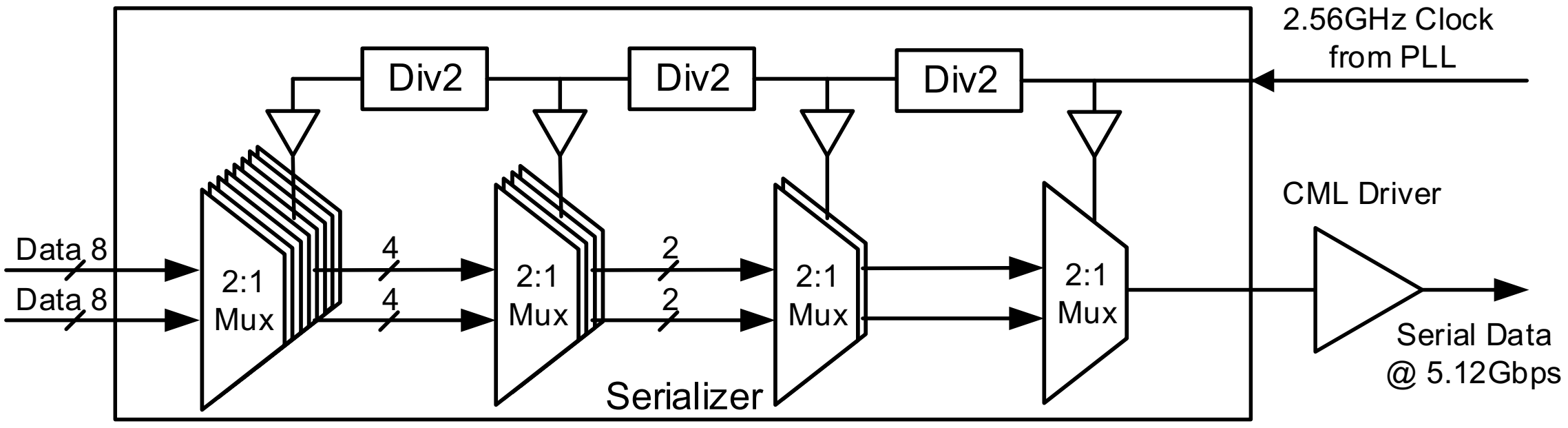

**Figure 5:** The block diagram of the serializer.

## 2.2 The serializers

The block diagram of the serializer is shown in figure 5. LOCx2 consists of two 16:1 serializers, each operating at 5.12 Gbps. Each serializer has 16-bit parallel data inputs in CMOS logic and a serial data output in CML logic. Each serializer consists of four stages of 2:1 multiplexers in a binary-tree structure. All 2:1 multiplexer are based on static CMOS D flip-flops for single-event effect immunity. Each serializer has a differential CML line driver with 50-Ω output impedance. The CML line driver is composed of five-stage CML differential amplifiers. Two serializers share one PLL, which provides the clock signals locked to the external reference clock.

## 2.3 The encoders

The ASIC uses a custom line code called LOCic [11]. The LOCic frame, defined as the data transmitted by each serializer in one LHC clock cycle, consists of an 8-bit frame header, the 112-bit payload and an 8-bit frame trailer. The total length of each LOCic frame is 128 bits. The frame header includes a 4-bit fixed pattern "1010" and 4-bit encoded Bunch-Crossing Identification (BCID) information based on two Pseudo-Random Binary Sequences (PRBSs). The payload includes the digitized data of eight analog signals and the calibration data. The 8-bit frame trailer is an 8-bit Cyclic Redundant Checking (CRC) code of the payload. The payload is scrambled before being transmitted, whereas neither the frame header nor the frame trailer is scrambled. The encoder overhead is 16/112 or 14.3%.

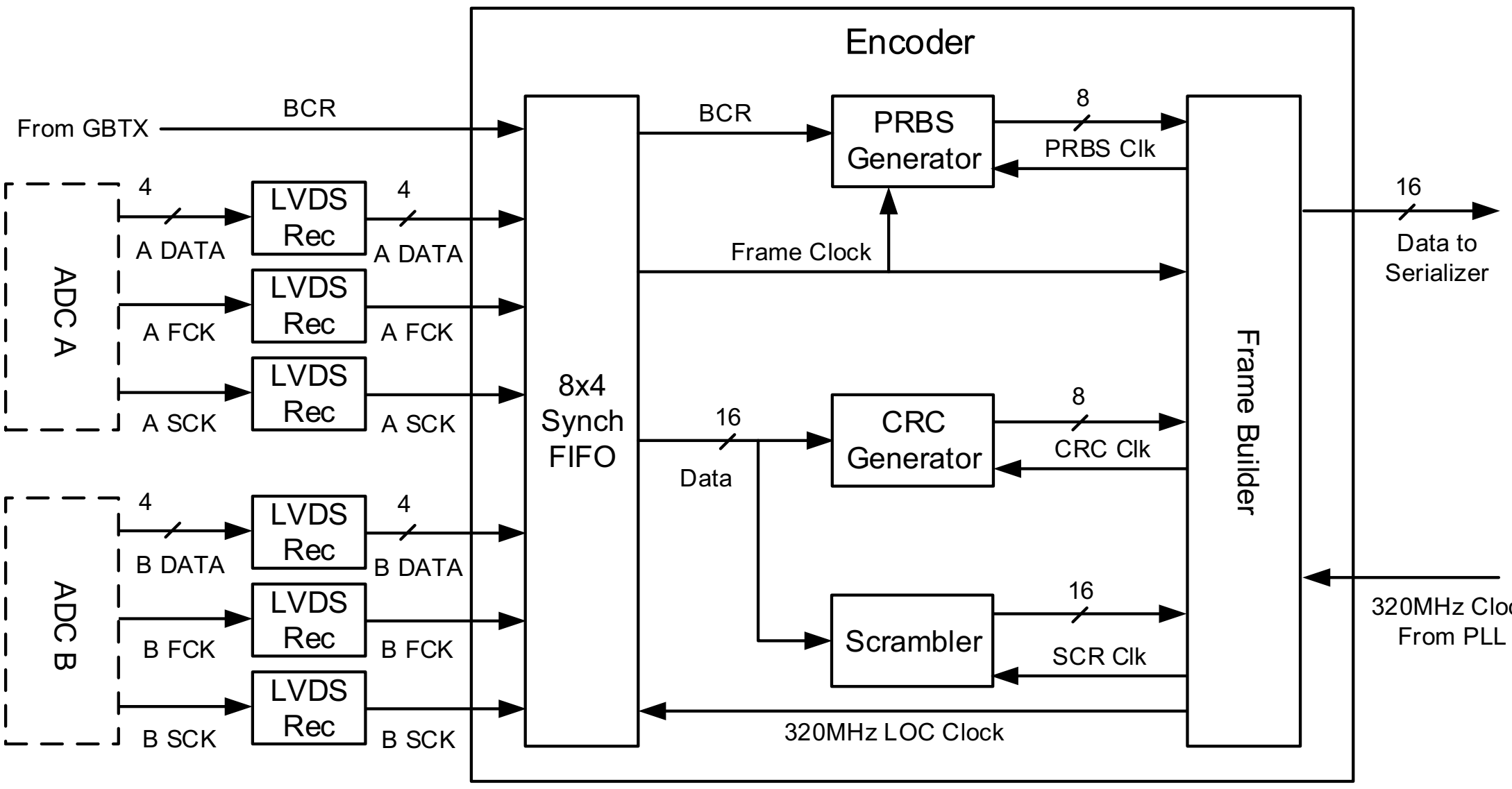

**Figure 6:** The block diagram of the LOCic encoder.

Figure 6 shows the block diagram of the LOCic encoder. The encoder is composed of a synchronous First-In-First-Out (FIFO) buffer, a PRBS generator, a CRC generator, a scrambler, and a frame builder. The PRBS generator generates the frame header. The CRC generator generates the CRC code of the payload. The scrambler scrambles the payload. The framer builder constructs a frame based on the framer header, the scrambled data, and the frame trailer. To reduce the latency and hardware cost, the encoder avoids from orientating the input data while adding an extra frame header and an extra frame trailer in each frame. In order to reduce the latency, the encoder is optimized to operate at 320 MHz, the highest operational frequency with reasonable design margins. Since such a speed cannot be achieved in the standard digital design procedure, the encoder is custom designed and laid out manually.

## 3. The test and measurement results of LOCx2

The block diagram of LOCx2 test setup is shown in figure 7. A clock board (Si5338 Evaluation Board Kit produced by Silicon Labs) generates three synchronized differential clock signals, two at 40MHz and one at 320 MHz, respectively. One 40-MHz clock is the reference input clock of LOCx2. The other 40-MHz clock provides the trigger for a 50-GSample/s real-time oscilloscope (Model DSA 72004 produced by Tektronix). The 320-MHz clock provides the reference input clock for a Xilinx Kintex-7 FPGA KC705 Evaluation Kit (Part number EK-K7-KC705-G produced by Xilinx). The serial-data outputs of LOCx2 are sent to either the high-speed real-time oscilloscope or the FPGA. The oscilloscope is used to observe eye diagrams and measure the jitter. In addition to the two 5.12-Gbps serial output data, each LOCx2 also has a 640-MHz test clock output to evaluate the PLL performance. We use an $I^2C$ master (Model USB-8451, Part No. 779553-01, produced by National Instruments) to configure the LOCx2 in the test. A picture of the test setup is shown in figure 8.

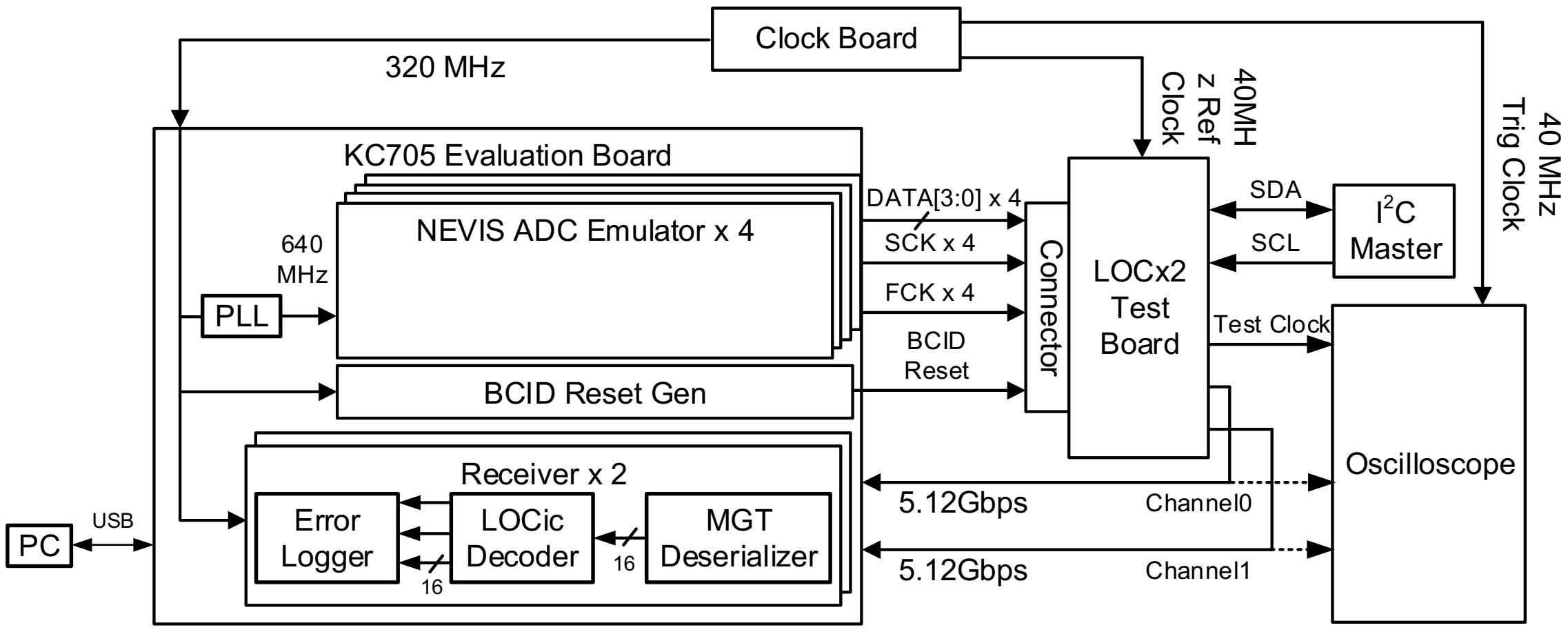

**Figure 7:** The block diagram of test setup.

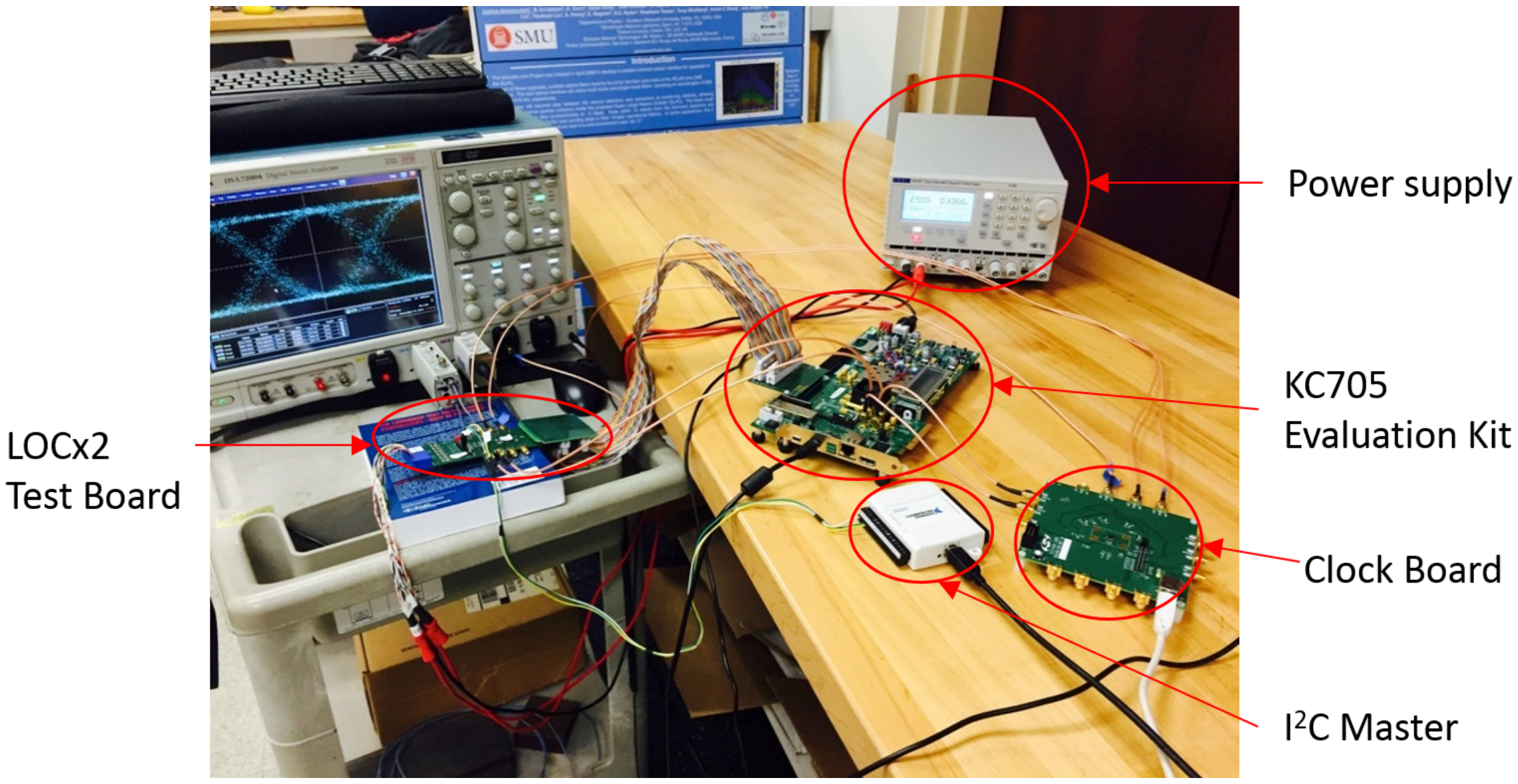

**Figure 8:** A picture of the test setup.

The FPGA implements four ADC emulators, a BCR generator, and two link receivers. Each ADC emulator generates PRBS data, a serial data clock (SCK), and a frame clock (FCK).

The FPGA also generates a BCR signal, which lasts for one cycle and repeats every 3564 cycles of the 40-MHz clock. The link receiver includes the deserializer of a Multiple-Gigabit Transceiver (MGT), a LOCic decoder, and an error logger. The deserializer converts the 5.12-Gbps serial data stream into 16-bit parallel data. Then the LOCic decoder recovers the original ADC data and checks if there are errors in the recovered data. The error logger records error types and time stamps. The error logger status is monitored on a personal computer via the Xilinx ChipScope Pro tool.

### 3.1 The serializer and PLL test

The performance of the LC-PLL was measured via the output test clock signal. The measured tuning range of the PLL is from 2.0 to 3.1 GHz. The random jitter of the output test clock signal is less than 1 ps (RMS).

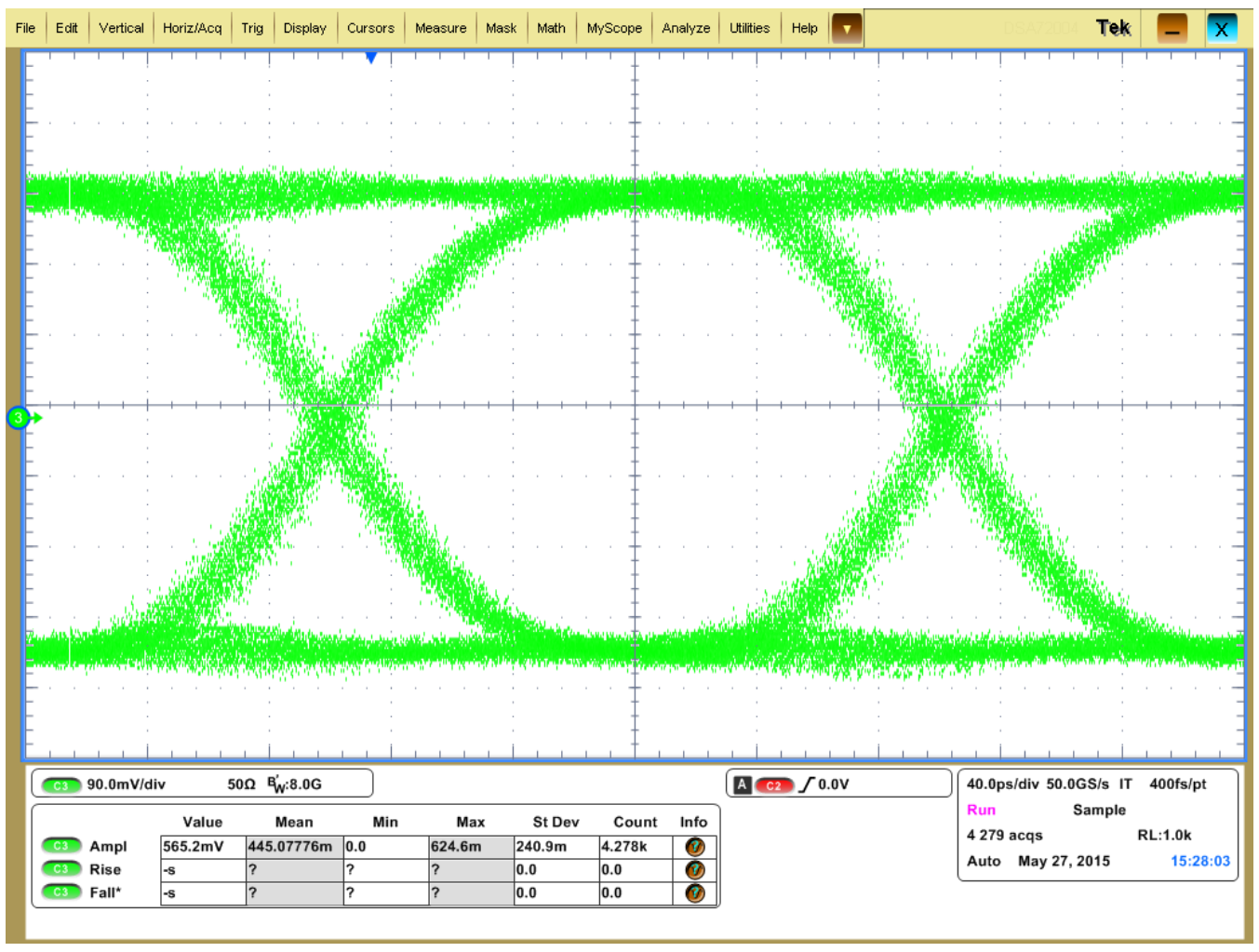

**Figure 9:** An eye diagram of LOCx2 at 5.12 Gbps.

The eye diagram of LOCx2 at 5.12 Gbps is shown in figure 9. The rise time and the fall time are about 70 ps and 64 ps, respectively. At the data rate of 5.12 Gbps, the deterministic jitter is 28 ps (peak-peak) and random jitter is 1.5 ps (RMS), corresponding to total jitter of less than 42 ps (peak-peak) at the bit error rate of $10^{-12}$. The typical differential amplitude of the output is 445 mV (peak-peak).

### 3.2 The encoder function test and link latency measurement

The LOCic encoder was tested. BCID information was completely recovered in the decoder and the CRC code regenerated from the received data was consistent with the received CRC code. Moreover, the received payload were consistent with that generated in the ADC emulator. No error was observed during the test for nine minutes, meaning that the bit error rate was less than $10^{-12}$ at the confidence level of 95%.

The latency of the whole link, including LOCx2 and the receiver implemented in an FPGA, not including optical transceiver and the optical fiber, was measured. The latency of the whole link is no more than 75.3 ns, meeting the design goal. The latency contribution of LOCx2 is less than 27.2 ns. The latency variation is due to the phase uncertainty of the internal clocks of the FIFO and the deserializer after each power cycle. The latency variation is 6.25 ns, much less than an LHC clock cycle. When the data are latched with a 40-MHz LHC clock on the receiver side and then sent to the trigger system, the latency is still fixed.

### 3.3 Power consumption

At room temperature and 5.12 Gbps, the power consumption of LOCx2 was measured to be about 843 mW, which is consistent with the simulated value and meets the design goal. The power consumptions of all functional blocks are listed in Table 1.

**Table 1: The power consumption breakout of all functional blocks**

| Functional blocks | Power consumption (mW) |
|---:|---:|
| LC-PLL | 52.5 |
| LOCic | 192.5 |
| Serializer | 350.0 |
| CML Driver | 150.0 |
| Clock Buffers | 22.5 |
| SLVS Receivers | 75.0 |
| Total | 842.5 |

## 4. Conclusion

We present the design and test results of LOCx2, a serializer ASIC for the ATLAS Liquid Argon Calorimeter trigger upgrade. LOCx2 consists of two serializers and each serializer encodes ADC data with an overhead of 14.3% and transmits serial data at 5.12 Gbps with a latency of less than 27.2 ns. The power consumption of LOCx2 is about 843 mW.

## Acknowledgments

This work is supported by US-ATLAS R&D program for the upgrade of the LHC, the US Department of Energy Grant DE-FG02-04ER1299, and the National Natural Science Foundation of China under Grant No. U1232206 and 11220101005. We are grateful to Drs. Sandro Bonacini and Paulo Moreira of CERN for their help in the $I^2C$ interface design, Mrs. Jee Libres and Nicolas Moses of VLISP Technologies, Inc. for their help in ASIC packaging, Drs. Hucheng Chen and Hao Xu of Brookhaven National Laboratory, Dr. Nicolas Dumont Dayot of LAPP, and Dr. Bernard Dinkespiler of CPPM for beneficial discussions in the encoder/decoder implementation and system integration.